\documentstyle[aps,prl,twocolumn]{revtex}
\begin{document}

\draft

\title{Oscillations of the superconducting order parameter in a ferromagnet}
\author{T. Kontos, M. Aprili, J. Lesueur and X. Grison}
\address{CSNSM-CNRS, University Paris-Sud, 91405 Orsay Cedex, France}
\date{\today}

\maketitle

\begin{abstract}
Planar tunneling spectroscopy reveals damped oscillations of the superconducting 
order parameter induced into a ferromagnetic thin film by the proximity effect. 
The oscillations are due to the finite momentum transfer provided to Cooper 
pairs by the splitting of the spin up and down bands in the ferromagnet. As a 
consequence, for negative values of the superconducting order parameter the 
tunneling spectra are capsized ("$\pi$-state"). The oscillations' damping
and period are set by the same length scale, which depends on the spin 
polarization.\\
\end{abstract}
\pacs{74.50.+r}

\narrowtext

The quantum character of superconductivity arises from the existence of phase 
coherence in the electron condensate. In conventional superconductors, where 
pairing is provided by the exchange of virtual phonons, the phase is a constant. 
On the other hand, phase sensitive experiments \cite{vah95} in high temperature 
superconductors have shown that the wavefunction of Cooper pairs with 
perpendicular quasiparticle momenta presents a $\pi$-phase shift suggesting 
unconventional pairing. Here, we show that a $\pi$-phase shift
 can also occur in the order parameter of conventional superconductors when 
superconducting correlations coexist with ferromagnetic order.

More than 30 years ago, Fulde and Ferrel \cite{ful64}, and Larkin and 
Ovchinnikov \cite{lar65} (FFLO), showed independently that the superconducting 
order parameter may be modulated in real space by an exchange field. A Cooper 
pair, in the singlet state, acquires a finite momentum $Q=2E_{ex}/ \hbar v_F$, 
where $2E_{ex}$ is the exchange energy corresponding to the difference in energy 
between the spin-up and spin-down bands, and $v_F$ the Fermi velocity. The 
superconducting phase grows linearly with the spatial coordinate $x$, $\varphi=Q 
\cdot x$ and a $\pi$-phase shift was expected for translations of $\Delta x 
\approx hv_F/4E_{ex}$. Unlike high temperature superconductors where
 $\varphi$ is a 2$\pi$-multiple of 0 and $\pi$, in the FFLO-state $\varphi$ 
varies continuously.

The FFLO-state only occupies a tiny part of the superconducting phase diagram 
close to the normal state \cite{ful64}. The fragility of singlet 
superconductivity in a finite exchange field that removes the degeneracy of the 
ground state with respect to the spin degrees of freedom makes its experimental 
evidence still an open question. In homogeneous superconductors the normal state 
is recovered when $E_{ex}> \sqrt{2}/2 \Delta_s$ (Clogston criterion) 
\cite{clo62}, where $\Delta_s$ is the superconducting energy gap. The situation 
is more favorable if Cooper pairs are injected from a superconductor into a 
ferromagnet F by the proximity effect. Assuming that the superconductor is 
weakly affected by the exchange field, superconducting correlations persist in F 
even for exchange energies much higher than $\Delta_s$. The physical reason is 
that
 Cooper pairs are not instantaneously broken when they penetrate into the 
ferromagnet. They survive for a time corresponding to a traveled length on the 
order of $\xi_F=\hbar v_F/2E_{ex}=1/Q$, the coherence length scale in F 
\cite{dem97}, which is independent on the energy gap. The breakdown of the 
Clogston criterion turns out to be very significant since $E_{ex}$ is typically 
at least two orders of magnitude larger than $\Delta_s$.

When a Cooper pair moves into a ferromagnet, the phase shift produces 
oscillations of the real part of the superconducting order parameter on a length 
scale given by $\xi_F$ \cite{dem97}, as shown in Fig.\ \ref{fig1}a \cite{note1}. 
However this artificially generated FFLO-state vanishes on the same length 
scale, which is typically of the order of a few $nm$. Unlike bulk 
superconductors where the gap equation allows the FFLO-state to
 occur only for exchange fields close to the critical field, this state exists 
for any exchange energy. Furthermore, the oscillating behavior of the order 
parameter in F is only due to the phase factor, while the number of Cooper pairs 
decays monotonically. We shall call the states corresponding to a positive sign 
of the real part of the order parameter the "$0$-state" and those
 corresponding to a negative sign of the order parameter the "$\pi$-state".

An induced superconducting order parameter in F modifies the quasiparticle 
density of state (DOS). In the "$\pi$-state", i.e. when the thickness of the 
ferromagnet is larger than $\xi_F$, the features in the superconducting DOS are 
reversed with respect to the normal state (see inset a of Fig.\ \ref{fig1}a). 
This can be explained considering the microscopic mechanism that allows 
superconducting correlations to propagate into F, i.e., Andreev
 reflections \cite{and64}. The process is illustrated in Fig.\ \ref{fig1}b using 
the energy-momentum dispersion law of the normal metal: an incoming electron in 
a normal metal N with energy lower than $\Delta_s$ from the Fermi level, is 
reflected into a hole at the S/N interface. The incoming electron and the 
outgoing hole accumulate a phase difference ($\varphi=\Delta p\cdot x$) 
depending on their traveled distance, $x$, and on the difference
 between their momenta, $\Delta p$. Note that $\Delta p$ is a function of the 
quasiparticle energy. If the normal layer is very thin, the phase difference is 
small and, roughly speaking, the DOS in N is close to that of the Cooper pair 
reservoir. The situation is strongly modified if the normal
 metal is ferromagnetic. As Andreev reflections invert spin-up into spin-down 
quasiparticles and vice-versa \cite{jon95,sou98}, the total momentum difference 
includes the spin-splitting of the conduction band: $\Delta p_F=\Delta p+Q$ (see 
Fig.\ \ref{fig1}c). If the exchange energy is much larger than the energy gap, 
which is usually the case, $\Delta p_F \approx Q$ and the phase difference 
between the electron and hole wavefunction is almost energy independent.
 The DOS is modified in a thin layer on the order of $\xi_F$. In particular, the 
interference between the electron and hole wavefunction produces an oscillating 
term in the superconducting DOS with period $\approx x E_{ex} /\hbar v_{F}$. A 
phase-induced oscillating term in the superconducting DOS is a natural 
consequence of the proximity effect. It has already been observed in the past 
either in the clean \cite{row66} or the dirty \cite{gue96} limits. However, 
oscillations usually appear as a function of energy since, in a normal metal, 
the phase is energy dependent. Differently, in a ferromagnet the oscillations 
turn the overall energy-dependent DOS up side down with respect to the normal 
state.

We measure the superconducting DOS by tunneling spectroscopy. The conductance 
vs. bias of a tunnel junction between a superconductor and a normal metal probes 
the energy-dependent quasi-particle DOS of the superconductor, convoluted by the 
thermal broadening. The DOS is obtained after normalizing the spectra by the 
background conductance measured when both electrodes are in the normal state. 
Planar junctions provide unsurpassed energy resolution and even more 
importantly, in our case, large magnitude resolution. We fabricated planar 
junctions completely "in-situ" by thin film deposition in a typical base 
pressure of $10^{-9}$ Torr. An aluminum layer is deposited on a silicon 
substrate and then oxidized in an oxygen glow discharge (P=$8~10^{-2}$ mbar) 
resulting in high quality tunnel barriers. The junction area 100$\mu$m x 
100$\mu$m is defined by evaporating 500 \AA~of insulator (SiO) through shadow 
masks. A thin layer of $Pd_{1-x}Ni_x$ (hereafter called PdNi) backed by 500 \AA~ 
of Nb ($T_c$=8.8 K) is deposited just after oxidation defining a four-terminal 
cross-junction geometry \cite{les97}. The Nb and the PdNi respectively provide 
the Cooper pair reservoir and the ferromagnetic thin film. The mean free path in 
the PdNi thin films is limited by surface scattering. The Ni concentration is 
kept on the order of 10 \% and measured after fabrication by Rutherford 
Back scattering Spectrometry (RBS). The junction resistance is typically between 
$50 \Omega$ and 1K$\Omega$ while the interface resistance between PdNi and Nb is 
$\approx 10^{-5} \Omega$. The bias-dependence of the tunneling conductance is 
measured using a standard AC-modulation technique.  Combining Lock-in detection 
with an ultra-low noise
 DC/AC mixer, we can directly resolve structures in the DOS as small as $10^{-
4}$ of the background conductance.

Ferromagnetic order in PdNi alloys results from indirect exchange between the Ni 
magnetic moments provided by the large spin susceptibility of Pd \cite{mur74}. 
At low Ni concentrations, the total magnetic moment is mainly due to the spin 
polarized electrons of the host at the Fermi level \cite{tak65}. Long-range 
itinerant ferromagnetism provides an almost ideal system where Cooper pairs are 
suddenly polarized when they enter into the ferromagnet. The main advantage of 
using a ferromagnetic alloy, instead of pure Ni, for instance, is that the 
exchange energy can be kept suitably small. $E_{ex}$ may be estimated from the 
magnetization $M \approx \mu_B \chi E_{ex}$, where $\mu_B$ is the Bohr magneton 
and $\chi$ the host susceptibility. In PdNi alloys with 10 \% of Ni, $E_{ex}$ is 
of the order of $10 meV$ \cite{bei75} resulting in $\xi_F \approx$ 50 \AA, which 
corresponds to an order of magnitude increase with respect to pure ferromagnetic 
elements such as Fe, Ni or Co. This coherence length is accessible to standard 
thin film technology. Of course, decreasing the Ni concentration closer to the 
paramagnetic-ferromagnetic transition would further increase the penetration 
length of Cooper pair into the ferromagnet. However, we observed that lowering 
the Ni concentration results in a reduced magnetic homogeneity.

In Fig.\ \ref{fig2}a the superconducting DOS at T=300 mK is presented for two 
different thickness of PdNi. The Al counter-electrode is driven into the normal 
state by applying a magnetic field of 100 Gauss perpendicular to the film 
\cite{note2} For the thinner ferromagnetic layer (50\AA) the phase factor is 
positive ("0-state") and the DOS displays a maximum at the Nb gap edge and a 
minimum at the Fermi level set to zero in our spectra. As a result of the finite 
interface resistance between PdNi and Nb, the pair amplitude is small, 
corresponding to a few per cent difference from the background conductance. To 
stress that in our geometry the relevant energy scale for the proximity effect 
is the Nb gap energy $\Delta_{Nb}$=1.40$meV$, the DOS of Nb measured in a 
junction without PdNi is also plotted on the r.h.s. of Fig.\ \ref{fig2}a. 
Increasing the thickness of the ferromagnetic layer (75\AA), the phase factor 
becomes negative ("$\pi$-state") and the DOS is flipped with respect to the 
normal state. When both electrodes are in the superconducting state the 
structures are amplified by the Al BCS singularity and shifted in energy by the 
aluminium gap as expected for elastic tunneling. 

A check on the magnetic properties of F is shown in Fig.\ \ref{fig2}b and Fig.\ 
\ref{fig2}c, which presents the normalized Hall resistivity, $\rho_{Hall} 
/\rho^2$, vs. applied field, of the 50 and 75 \AA~thick~PdNi layers 
respectively, corresponding to the "0 and $\pi$-state" measured by tunneling 
spectroscopy. The Hall resistivity is sensitive to magnetic scattering 
through the spin-orbit coupling and provides a suitable probe of weak magnetic 
moments in thin films \cite{ber78}. In ferromagnetic materials scattering by 
defects produces a net asymmetry in the transverse current density that is 
compensated, at equilibrium, by the anomalous Hall field \cite{hur72}. The Hall 
resistivity shows a fast variation at low magnetic field when the magnetic 
domains order and a linear dependence at higher field corresponding to the 
ordinary Hall effect. As the anomalous Hall effect is proportional to the 
magnetization and to the square of $\rho$, the film resistivity,
the extrapolation of $\rho_{Hall} /\rho^2$ at zero field is directly 
proportional to the saturation magnetization \cite{ber80}. 
Complementary measurements by MOKE (Magneto-Optical-Kerr-Effect) on junctions 
presenting the same structure also show ferromagnetic ordering with a typical 
coercive field, $H_c$, of 1500 Gauss close to that measured by the anomalous 
Hall effect ($H_c$= 1200 Gauss). Finally, from the direct measurement of the 
saturation magnetization by SQUID we can extract the exchange energy and hence 
verify the estimated coherence length in the ferromagnet. We obtain $M=0.21 
\mu_B$ which gives $E_{ex}=15meV$ and $\xi_F = 45$ \AA.

Increasing the thickness of the ferromagnetic layer, i.e. for $x >> \xi_F$, the 
proximity effect disappears and the normalized tunneling conductance becomes 
equal to unity. The DOS at zero energy, $N(0)$, as a function of $x$ is simply 
found from the spatial dependence of the order parameter, being $N(0)=\Re e ( 
\sqrt{1-\Psi^{2}} )$. In Fig.\ \ref{fig3} we present the DOS at zero energy vs. 
$x/\xi_F$. We observe quantitative agreement with the $cos (x \sqrt{2}/ \xi_F ) 
e^{-x \sqrt{2}/ \xi_F}$ dependence of the order parameter \cite{dem97,note1} 
presented in Fig.\ \ref{fig1}a. Here $\xi_F$ is the coherence length in the 
dirty limit and is obtained by measuring the exchange energy from the saturation 
magnetization as indicated above. The only fitting parameter is the finite 
interface transparency, $\gamma_B=0.035$, which accounts for the reduced pair 
amplitude in F. A shift of 15 \AA~in the thickness of the ferromagnetic layer 
was also included to account for interface roughness or interdiffusion as shown 
by X-ray reflectivity measurements \cite{coc98} 

Our results show that the superconducting order parameter induced into a 
ferromagnet by proximity effect oscillates with a period given by the exchange 
energy. They suggest that S/F nanostructures offer a unique way to investigate 
the interplay between superconductivity and magnetic order since they do not 
require comparable energy scales. Furthermore they indicate that the proximity 
effect may indeed be used to fabricate Josephson junctions with a $\pi$-phase 
shift, as recently proposed \cite{buz92}.

We are indebted to P. Veillet, P. Monod and J. Ferr\'{e} for the SQUID and MOKE 
magnetic characterization. We acknowledge W. Guichard and P. Gandit for 
measuring the Nb/PdNi interface resistance. We also thank D. Esteve, H. Pothier, 
W. Belzig , Yu. Nazarov, A. Buzdin,  J-P. Torre, O. Bourgeois and H. Bernas for 
useful discussions.

\begin{figure}
\caption{ a) Exponentially damped oscillations of the real part of the 
superconducting order parameter induced into a ferromagnetic material by 
proximity effect. The space coordinate $x$ denotes the distance from the 
superconductor/ferromagnet interface. The period of the oscillations is set by 
the coherence length $\xi_F$. "$0$-state" and "$\pi$-state" correspond to a 
positive and negative sign of the real part of the order parameter, 
respectively. For the sake of simplicity, the superconductor is assumed 
unaffected by the exchange field of the ferromagnet F. Inset : superconducting 
density of states at zero temperature in the "$0$ and $\pi$" state for an 
exchange energy $E_{ex}$ much larger than the energy gap, $\Delta_s$. The 
characteristic reversed shape in the "$\pi$-state" is a consequence of the order 
parameter oscillations. b) Schematic of the Andreev reflection process: an 
electron in the normal metal with momentum, $k_+$, is elastically reflected as a 
hole, $k_-$, at the superconductor/normal metal interface (S/N). c) If N is spin 
polarized the momentum shift, $\Delta p_F$, is dominated by the spin-splitting 
of the up and down bands.}
\label{fig1}
\end{figure}

\begin{figure}
\caption{ a) Differential conductance vs. bias for two 
Al/Al$_{2}$O$_{3}$/PdNi/Nb tunnel junctions corresponding to two different 
thickness (50 \AA~and 75 \AA) of PdNi . The spectra have been taken at T=300 mK 
and H=100 Gauss and normalized by the normal state conductance obtained applying 
a magnetic field higher than the Nb critical field. The tunneling spectra show 
the "$0$" and "$\pi$" state shape expected from Fig.1a when the thickness of the 
ferromagnetic layer is respectively smaller or larger than $\xi_F$. Note that 
the induced superconducting density of states (DOS) is small. The normalized 
conductance for a tunnel junction without PdNi is also reported on the r.h.s. 
The field dependence of the normalized Hall resistivity at T=1.5 K for the same 
PdNi films as in the tunnel junctions corresponding to the "$0$-state" (50 \AA) 
and to the "$\pi$-state" (75\AA) is shown in Fig. 3 b) and c) respectively. Long 
range magnetic order leads to saturation of the magnetic domains and field-
induced hysteresis.}
\label{fig2}
\end{figure}

\begin{figure}
\caption{Tunneling conductance at zero energy vs. the PdNi thickness normalized 
by the coherence length $\xi_F$. The data taken at T=300 mK and H=100 G, are 
shown as solid symbols. The theoretical curve (dotted line) obtained solving the 
Usadel equations in the presence of an exchange field takes into account a 
finite interface resistance as fitting parameter. The dashed line denotes the 
transition from the "$0$-state" to the "$\pi$-state".}
\label{fig3}
\end{figure}

\end{document}